# Non-obstructive intracellular nanolasers

Alasdair H. Fikouras[1], Marcel Schubert[1], Markus Karl[1], Jothi D. Kumar[1], Simon J. Powis[2], Andrea di Falco[1,*], Malte C. Gather[1,*]

[1] *SUPA, School of Physics and Astronomy, University of St Andrews, St Andrews, U.K.*

[2] *School of Medicine, University of St Andrews, St Andrews, U.K.*

*email: mcg6@st-andrews.ac.uk, adf10@st-andrews.ac.uk

**Nanophotonic objects like plasmonic nanoparticles and colloidal quantum dots can complement the functionality of molecular dyes in biomedical optics[1-7]. However, their operation is usually governed by spontaneous processes, which results in broad spectral features and limited signal-to-noise ratio, thus restricting opportunities for spectral multiplexing and sensing. Lasers provide the ultimate spectral definition and background suppression, and their integration with cells has recently been demonstrated[8-12]. However, laser size and threshold remain problematic. Here, we report on the design, high-throughput fabrication and intracellular integration of semiconductor nanodisk lasers. By exploiting the large optical gain and high refractive index of GaInP/AlGaInP quantum wells, we obtain lasers with volumes 1000-fold smaller than the eukaryotic nucleus ($V_{laser}$<0.1 μm³), lasing thresholds 500-fold below the pulse energies typically used in two-photon microscopy ($E_{th}$≈0.13 pJ), and excellent spectral stability (<50 pm wavelength shift). Multiplexed labelling with these lasers allows cells-tracking through micro-pores, thus providing a powerful tool to study cell migration and cancer invasion.**

Lasers embedded in tissue or even in the cytoplasm of a cell have recently been used for high-density optical barcoding of cells and to perform local optical sensing, and have been suggested as non-linear probe for super-resolution imaging[13-15]. The lasers used so far occupy a substantial fraction of the cell volume (typical volume of eukaryotes, 1,000–10,000 μm³; volume of the nucleus, the largest stiff component in most eukaryotes, ≈100 μm³). To achieve lasing from whispering gallery mode (WGM)[16] resonances in fluorescently doped polystyrene spheres or oil droplets, a diameter >10 μm is required (volume, >500 μm³)[9,10]. Very recently, intracellular lasing from nanowires with linear rather than spherical form-factor has been demonstrated, with reduced volume (≤1 μm³) but with lengths in the range of 3–8 μm[11,12]. However, for many studies one will require sub-μm size in each dimension, e.g. to allow migration of cells through capillaries (typical diameter, 5 μm) and epithelial layers (pore and channel size in migration and nuclear rupture assays, 1.5 μm)[17].

Miniaturization of lasers, down to deep sub-wavelength dimensions, is an area of very active research in optical computing and communication and is widely regarded as one of the most promising avenues to address the ever-increasing demand for speed and bandwidth in data transmission and information processing[18-20]. However, the performance targets and trade-offs pertaining to miniaturization of intracellular lasers are considerably different: intracellular lasers operate in an aqueous environment which poses demands on their chemical stability and leads to reduced refractive index contrast between the laser material and its environment; the laser material, the resonator and the pump should have the least possible impact on cell physiology; to integrate intracellular lasers with other bioimaging technology platforms, they should operate in a spectral window already used for *in vivo* microscopy and where tissue scattering and absorption are low.

Figure 1a shows a schematic illustration of the intracellular lasers developed here. The lasers are formed by WGM disk resonators with minute volumes (~0.1 µm$^3$) and sub-micron diameters (down to 700 nm). The laser material is an epitaxially grown aluminium gallium indium phosphide (AlGaInP) multi-quantum-well structure that provides large optical gain[21,22]. This material platform has the additional benefit of being free of the highly toxic Arsenic often found in other III-V semiconductor lasers. We found that our disk lasers are readily introduced into a variety of cells via natural endocytosis, including e.g. into human T-cells, which are too small to engulf previously reported polymer-, oil- or nanowire-based intracellular lasers. The WGMs supported by each disk have low loss and showed lasing upon optical pumping of the disk with sub-pico-joule pulses of light. We show that the lasing wavelength is highly dependent on disk size, allowing unique labelling of large numbers of cells by introducing one or several nanodisk lasers of distinct diameter. Importantly, the nanodisk lasers introduced here are small enough to allow unobstructed migration of cells in confined environments.

In practice, the miniaturization of most laser resonators is hampered by the sharp increase in losses at small sizes; efficient lasers typically require resonators with $Q$ factors above $10^3$–$10^4$. For WGM resonators, the size-dependent, radiative component of the $Q$ factor depends on the resonator diameter and on the refractive index contrast with the environment (Fig. 1b). Previously reported intracellular lasers were based on microsphere WGM resonators made from materials with $n_{sphere} \approx 1.6$. Optical modelling shows that the low index contrast between these spheres and cells (typical $n_{cell} \approx 1.37$)[23] requires a minimum sphere diameter of 10 µm to reach $Q = 10^3$. By contrast, our AlGaInP quantum-well structures provide an average refractive index of $n_{AlGaInP} \approx 3.6$ and we therefore expect to achieve sufficient $Q$ for lasing in cells with resonator diameters well below 1 µm.

A further advantage of producing intracellular lasers from a semiconductor quantum-well structure epitaxially grown onto a support wafer is the availability of well-established electron beam-based nanolithography processes to accurately control size and shape. To render these processes compatible with biological samples, we developed a combined under-etch and washing process to detach the nanodisks from their support wafer and transfer them at high yield into cell culture medium (Fig. 1c, Methods).

Figure 2a shows a scanning electron microscopy picture of our nanodisk lasers. In this example, the disk detachment process was interrupted just before completing the under-etch, leaving individual disks with diameters around 750 nm that rest on slender pillars. Using this configuration, we established that gradual minute changes in disk size across the wafer allow tuning of the lasing wavelength (Fig. 2b).

To confirm lasing action and characterize the laser performance, disks were transferred into a Petri dish filled with cell culture medium. Individual disks were excited by a pulsed diode-pumped solid-sated laser that was coupled to an optical microscope (pump wavelength, 473 nm; pulse duration, 1.5 ns; diameter of pump spot on sample, 4 µm). The emission from the disk was collected through the same objective, filtered from the pump light and passed to a digital camera and a spectrometer (spectral resolution, ≈0.2 nm FWHM). Figure 2c shows the intensity of light emitted by a representative nanodisk versus the peak intensity of the pump pulses, revealing a characteristic s-shape behaviour on the log-log plot, with a lasing threshold fluence of $E_{th}$ = 30 µJ/cm$^2$ (i.e. lasing started if ≥0.13 pJ pulse energy was incident on the nanodisk). Around the lasing threshold, a sharp spectral line emerged on top of the initially broad photoluminescence spectrum of the nanodisk and at higher pump intensities, the emission spectrum was dominated entirely by the lasing peak (signal-to-background ratio, 24 dB; FWHM of peak, 0.2 nm, i.e. limited by the resolution of the spectrometer). Our nanodisk lasers showed stable performance for 12 weeks when kept under physiological conditions (i.e., in cell culture medium and at 37°C). During 50 min of continuous laser operation under these conditions (i.e. >10$^5$ pump pulses, the longest we tested so far), the peak lasing wavelength fluctuated by less than 50 pm (Fig. 2e, determined by peak fitting to the lasing spectra).

A range of different cell types readily internalized our nanodisk lasers via natural phagocytosis (Fig. 3a). Importantly, internalization was also observed for T cells which are too small to engulf the previously reported cell lasers and which represent an important target for cell tagging due to their complex role in cancer progression and immunotherapies[24,25]. Uptake of nanodisk lasers was also observed into the soma of cultured primary neurons, which is typically not substantially larger than the cell nucleus it contains, again illustrating the importance of using sub-µm sized lasers for cell internalization. The presence of the nanodisk laser had no noticeable effect on cell behaviour and cell viability. Specifically, a cell viability / cell death assay showed 100% viability of disk-containing macrophages after 24 h (*n* =110 cells); cultures of primary cells containing disks remained viable for at least 2 weeks.

Nanodisks continued to produce narrow band laser light when inside a cell (Fig. 3b; FWHM, 0.2–0.25 nm). We followed a nanodisk laser inside a cell over 8 h and found that its intensity varied over time, presumably due to cellular motion. However, the lasing wavelength remained constant, i.e. the peak lasing wavelength changed by <80 pm (Fig. 3c). Lasing spectra also maintained high signal-to-background ratios of above 20 dB. During this experiment, the cell internalized a further nanodisk laser and the change in the refractive index from cell culture medium to inside the cell caused a clearly resolvable shift in lasing wavelength of 0.6 nm. Confocal microscopy performed on the cells that were fixed and stained at the end of the lasing experiment confirmed complete cellular internalization of the nanodisk lasers (Fig. 3d).

Due to their small size, each cell can internalize multiple nanodisk lasers without adding a heavy payload. The multiplexed emission spectrum of these lasers provides a highly characteristic optical barcode, thus allowing to uniquely label large numbers of cells. Figure 4 shows lasing spectra and microscopy images of cells with *N* = 1, 2, 3, and 6 lasers, and compares spectra recorded 1 h apart. The spectral shape of all lasers remained highly conserved. The emission from the *N* = 6 cell illustrates that even very closely spaced lasing peaks can be resolved separately due to the large signal-to-background ratio of the laser emission. As a conservative estimate, we assume that lasers can be clearly identified if their peak wavelengths are >0.4 nm apart (i.e. five-fold more than the maximum wavelength shift observed here). Given the wavelength tuning range of our nanodisk lasers (40 nm) and $N = 6$ disks per cell, one could thus uniquely tag >$10^9$ cells.

Finally, we demonstrate that cells tagged with nanodisks can migrate within spatially confined environments. For this, we adapted a widely used transwell migration assay and employed a nutrient gradient as driving force for migration through the pores (Fig. 5a). Using a confocal microscope that was integrated with the laser characterization setup, we observed multiple laser tagged cells that crossed onto the opposite side of the membrane (Fig. 5b). Nanodisk lasers on either side of the membrane emitted characteristic, narrowband lasing spectra (Fig. 5c).

The combination of low lasing threshold and bright narrowband emission within the detection range of silicon makes semiconductor nanodisk lasers introduced here an attractive candidate for a range of biological assays. Due to their small size and compact form factor, they are particularly attractive for applications requiring non-obstructive tagging of cells. Cell migration through pores and epithelial layers plays a crucial role, e.g. in cancer invasion and immune response[26]. Many cells migrate through pores with sizes down to ≈5 µm and under certain conditions cells unravel nuclear DNA to pass through openings as small as 1.5 µm[17,27]. Using our lasers to unambiguously track the trajectory of individual cells during these processes will provide important new insights, particularly when combined with single cell genomics and proteomics[28].

**Methods**

**Nanodisk fabrication.** Nanodisks were produced from an epitaxially grown quantum well structure located on a sacrificial AlGaAs layer and a sacrificial GaAs wafer (EPSRC National Centre for III-V Technologies, Sheffield). The specific layer structure used was GaAs (substrate), $Al_{0.77}Ga_{0.23}As$ (800 nm, sacrificial), InGaP (10 nm, protection), $(Al_{0.7}Ga_{0.3})_{0.51}In_{0.49}P$ (58 nm), 2x [$(Al_{0.5}Ga_{0.5})_{0.51}In_{0.49}P$ (10nm), $Ga_{0.41}In_{0.59}P$ (7 nm)], $(Al_{0.5}Ga_{0.5})_{0.51}In_{0.49}P$ (10nm), $(Al_{0.7}Ga_{0.3})_{0.51}In_{0.49}P$ (58 nm), InGaP (10 nm, protection). A 500 nm-thick layer of SU8 2000.5 resist (Microchem) was spun onto the substrate, patterned by electron beam lithography at 30KV (Raith E-line plus), and crosslinked on a hot plate at 90°C for 2 min. After development in Ethyl Lactate for 1 min, the sample was cured for 20 min at 180°C, and exposed to low pressure oxygen plasma for 1 min to fully clear the unexposed substrate. Samples were then immediately etched in a 2:5:100 $Br_2$:HBr:$H_2O$ solution for 10 s to define columns. Following complete removal of SU8 in oxygen plasma, samples were placed upside down in a plastic cell culture dish with 2.5% HF:$H_2O$ solution to detach disks directly into the dish. To eliminate residuals of the etchants before introducing cells, the content of the culture dish was neutralized with concentration and volume matched NaOH and then washed with DI water and phosphate buffered saline to remove the soluble NaF byproduct. Finally, dishes were filled with culture medium.

**Cell culture.** Cells were cultured in DMEM or $CO_2$ independent RPMI with 10 vol% fetal bovine serum (FBS, Thermo Fisher Scientific) and 1 vol% penicillin-streptomycin (PS, Thermo Fisher Scientific). To introduce nanodisk lasers into cells, cells were trypsinsed using trypsin-EDTA (0.25%, Thermo Fisher Scientific) and seeded into disk-containing culture dishes and incubated as usual. Cells were maintained in a humidified incubator at 37 °C and 5 % $CO_2$.

Primary macrophages and T cells were generated after ethical review and informed consent from blood of healthy donors. Primary neurons were obtained from C57 mouse pups (postnatal day 2–3) and cultured in Neurobasal-A supplemented with B-27 and 0.5 mM GlutaMax-I (Thermo Fisher Scientific). Animal procedures were in accordance with the United Kingdom Animals (Scientific Procedures) Act 1986.

For the confocal image shown in Fig. 3c, cells were fixed in paraformaldehyde (4%).

The cell viability / cell death assay used Calcein and propidium iodide nuclear stain according to the manufacturer's protocol (Sigma Aldrich). Statistics were collected from green and red channel epi fluorescence images, with blue and far red channels used for nuclear counting and nanodisk identification.

Transwell migration assays were performed using inserts with translucent PET membranes with 8 µm diameter pores (Greiner Bio-One). Inserts were mounted in coverslip-bottom Petri dishes using a custom holder that allowed *in situ* confocal sectioning of cells and disks with a high NA oil immersion objective. The top of membrane insert was seeded with eGFP expressing NIH 3T3 cells containing nanodisks in FBS depleted (0.1 vol%) medium. A nutrient gradient was established by filling the bottom compartment with FBS rich medium (30 vol%). Cells were incubated overnight under standard culture conditions prior to imaging.

**Optical setup.** Nanodisk containing cells were characterized on an inverted optical microscope, equipped with epi fluorescence, differential interference contrast (DIC) and a laser scanning confocal scanhead (Nikon). The output from a pulsed diode pumped solid state laser that was set to 100 Hz repetition rate (Alphalas) was coupled into the microscope via a dichroic filter and passed to the sample through either a 60x oil immersion or a 40x long working distance objective. Emission from disks was collected by the same objective, separated from the pump light by the dichroic and passed to the camera port of the microscope. The image of the sample was then relayed to a 300 mm spectrograph (Andor) and a cooled sCMOS camera (Hamamatsu). A green bandpass filter placed in the dia illumination path of the microscope, a removable dichroic beam splitter and additional band pass filters at the spectrograph and camera allowed simultaneous recording of the spectral output of nanodisks and DIC imaging of cells. The setup also allowed combining investigation of lasers with recording of epi fluorescence and confocal microscopy images, without a need to move the sample. A motorized xy stage with trigger interface (Prior Scientific) enabled automated sequential excitation of a large number of disks across an extended area of the sample. Long-term live cell imaging was rendered possible through use of an on-stage incubator system (Bioscience Tools).

**Modelling.** Radiative $Q$ factors of disks with different radii and in different media were obtained via finite element modelling (COMSOL Multiphysics), using the perfectly matched layer (PML) approach to obtain the complex eigenfrequencies for modes within the optical gain region[29,30]. PML thickness, distance and growth factor were optimized to avoid numerical instabilities. Disks were modelled as solid isotropic structures with a uniform, isotropic refractive index of 3.6. Radiative $Q$ factors of spheres were modelled using a semiclassical (WKB) approximation for the Riccati-Bessel radial solutions[16].

**References**

1  Aioub, M. & El-Sayed, M. A. A Real-Time Surface Enhanced Raman Spectroscopy Study of Plasmonic Photothermal Cell Death Using Targeted Gold Nanoparticles. *J. Am. Chem. Soc.* **138**, 1258-1264 (2016).

2  Köker, T. *et al.* Cellular imaging by targeted assembly of hot-spot SERS and photoacoustic nanoprobes using split-fluorescent protein scaffolds. *Nature Communications* **9**, 607 (2018).

3  Levy, E. S. *et al.* Energy-Looping Nanoparticles: Harnessing Excited-State Absorption for Deep-Tissue Imaging. *ACS Nano* **10**, 8423–8433 (2016).

4  Zhao, T. *et al.* A transistor-like pH nanoprobe for tumour detection and image-guided surgery. *Nature Biomedical Engineering* **1**, 0006 (2016).

5  Shambat, G. *et al.* Single-cell photonic nanocavity probes. *Nano Letters* **13**, 4999-5005 (2013).

6  Wang, D., Zhao, X. & Gu, Z. Advanced optoelectronic nanodevices and nanomaterials for sensing inside single living cell. *Optics Communications* **395**, 3-15 (2017).

7  Zhao, Y. *et al.* Microfluidic Synthesis of Barcode Particles for Multiplex Assays. *Small* **11**, 151-174 (2014).

8  Gather, M. C. & Yun, S. H. Single-cell biological lasers. *Nature Photonics* **5**, 406-410 (2011).

9  Schubert, M. *et al.* Lasing within Live Cells Containing Intracellular Optical Microresonators for Barcode-Type Cell Tagging and Tracking. *Nano Letters* **15**, 5647-5652 (2015).

10  Humar, M. & Yun, S. H. Intracellular microlasers. *Nature Photonics* **9**, 572-576 (2015).

11  Feng, C. *et al.* Organic-Nanowire-SiO2 Core-Shell Microlasers with Highly Polarized and Narrow Emissions for Biological Imaging. *ACS Appl. Mater. Interfaces* **9**, 7385-7391 (2017).

12  Wu, X. *et al.* Nanowire lasers as intracellular probes. *Nanoscale* **10**, 9729-9735 (2018).

13  Schubert, M. *et al.* Lasing in Live Mitotic and Non-Phagocytic Cells by Efficient Delivery of Microresonators. *Scientific Reports* **7**, 40877 (2017).

14  Humar, M., Dobravec, A., Zhao, X. & Yun, S. H. Biomaterial microlasers implantable in the cornea, skin, and blood. *Optica* **4**, 1080-1085 (2017).

15  Cho, S., Humar, M., Martino, N. & Yun, S. H. Laser Particle Stimulated Emission Microscopy. *Physical Review Letters* **117**, 193902 (2016).

16  Chiasera, A. *et al.* Spherical whispering-gallery-mode microresonators. *Laser Photonics Reviews* **4**, 457-482 (2010).

17  Denais, C. M. *et al.* Nuclear envelope rupture and repair during cancer cell migration. *Science* **352**, 353-358 (2016).

18  Oulton, R. F. *et al.* Plasmon lasers at deep subwavelength scale. *Nature* **461**, 629–632 (2009).

19  Khajavikhan, M. *et al.* Thresholdless nanoscale coaxial lasers. *Nature* **482**, 204–207 (2012).

20  Hill, M. T. & Gather, M. C. Advances in small lasers. *Nature Photonics* **8**, 908-918 (2014).

21  Zhang, Z. *et al.* Visible submicron microdisk lasers. *Applied Physics Letters* **90**, 111119 (2008).

22  Qian, F. *et al.* Multi-quantum-well nanowire heterostructures for wavelength-controlled lasers. *Nature Materials* **7**, 701–706 (2008).

23  Liu, P. Y. *et al.* Cell refractive index for cell biology and disease diagnosis: past, present and future. *Lab Chip* **16**, 634-644 (2016).

24  De Simone, M. *et al.* Transcriptional Landscape of Human Tissue Lymphocytes Unveils Uniqueness of Tumor-Infiltrating T Regulatory Cells. *Immunity* **45**, 1135-1147 (2016).

25  Zheng, C. *et al.* Landscape of Infiltrating T Cells in Liver Cancer Revealed by Single-Cell Sequencing. *Cell* **169**, 1342-1356 (2017).

26  Au, S. H. *et al.* Clusters of circulating tumor cells traverse capillary-sized vessels. *Proc. Natl. Acad. Sci.* **113**, 4947-4952 (2016).

27  Raab, M. *et al.* ESCRT III repairs nuclear envelope ruptures during cell migration to limit DNA damage and cell death. *Science* **352**, 359-362 (2016).

28  Papalexi, E. & Satija, R. Single-cell RNA sequencing to explore immune cell heterogeneity. *Nature Reviews Immunology* **18**, 35-45 (2018).

29  Oxborrow, M. Traceable 2-D Finite-Element Simulation of the Whispering-Gallery Modes of Axisymmetric Electromagnetic Resonators. *IEEE Transactions on Microwave Theory and Techniques* **55**, 1209-1218 (2007).

30  Cheema, M. I. & Kirk, A. G. Implementation of the perfectly matched layer to determine the quality factor of axisymmetric resonators in COMSOL. *COMSOL Conference* (2010).

**Acknowledgements**

We thank Liam O'Faolain (CIT, Ireland) for fruitful initial discussion, Andrew Morton for support with neuronal culture, and Gareth Miles for kind provision of neuronal tissue samples. This research was financially supported by the European Research Council under the European Union's Horizon 2020 Framework Programme (FP/2014-2020)/ERC Grant Agreement No. 640012 (ABLASE), by EPSRC (EP/P030017/1, EP/L017008/1) and by the RS Macdonald Charitable Trust. AHF and MK acknowledge support through the EPSRC DTP (EP/M508214/1, EP/M506631/1). MS acknowledges funding by the European Commission (Marie Sklodowska-Curie Individual Fellowship, 659213) and the Royal Society (Dorothy Hodgkin Fellowship, DH160102).

**Author contributions**

AHF fabricated the nanodisks. AHF and MS characterized nanodisks and carried out the cell experiments. MK performed optical modelling. JDK optimized the transwell migration assay. SJP was responsible for cultures of primary macrophages and T cells. ADF and MCG supervised the project. MCG wrote the manuscript with input from all authors.

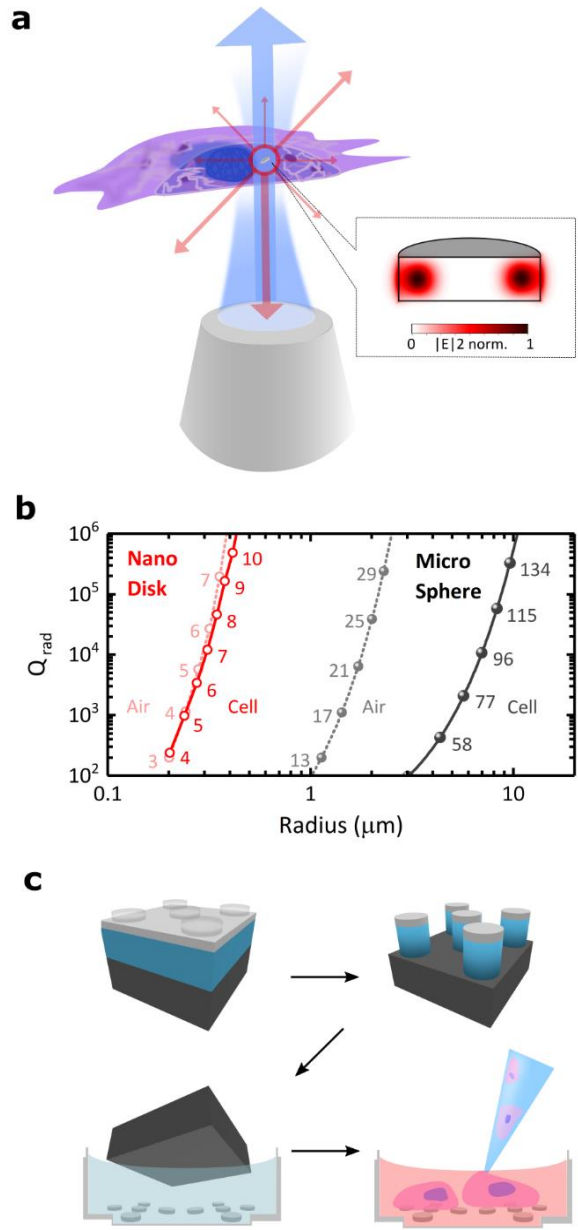

**Fig. 1 | Concept and modelling of intracellular nanodisk lasers. a**, Illustration of a semiconductor nanodisk laser internalized into a cell by natural endocytosis. The disk is optically pumped through a microscope objective (blue) with laser emission (red) collected by the same objective. Insert shows the calculated profile of the lowest radial order TE mode for a 750 nm diameter disk made of a GaInP/AlGaInP quantum well structure. **b**, Finite element modelling of the radiative $Q$-factor of lowest radial order TE modes in whispering gallery mode micro-resonators with different radii. Comparison between GaInP/AlGaInP nanodisks and polystyrene microspheres, both placed either in air or within a cell. Numbers next to each symbol indicate the angular quantum number of the corresponding mode. Lines are guides to the eye. Vacuum wavelength 680 nm. **c**, Schematic illustration of the nanolithography based fabrication of nanodisk lasers, the under-etch process for transfer into cell nutrient medium, and the subsequent culture of cells for disk internalization.

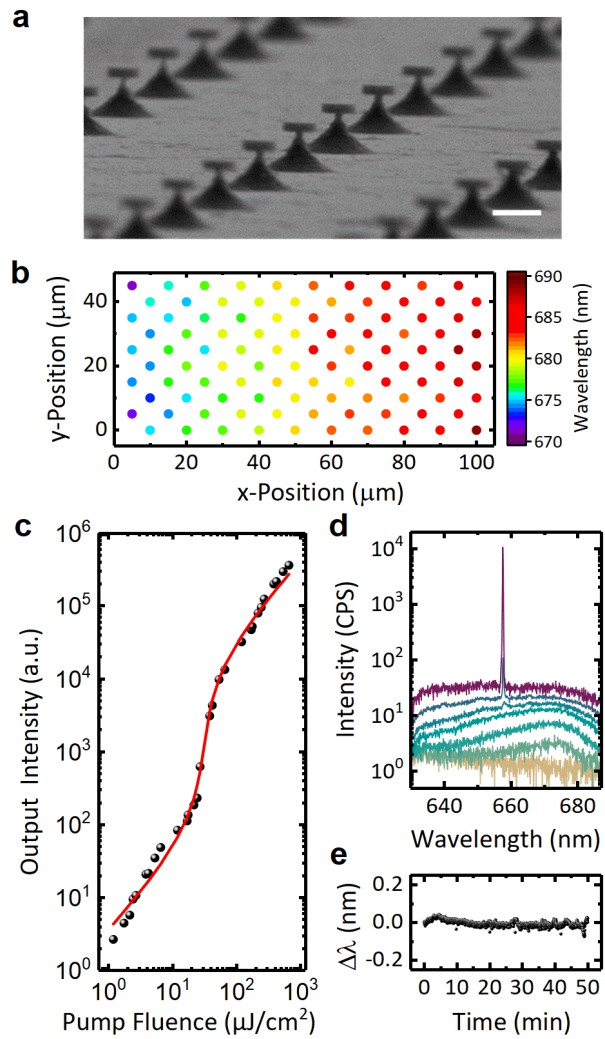

**Fig. 2 | Nanodisk laser characterisation. a**, Scanning electron microscopy of an array of as fabricated nanodisks on struts. Scale bar, 1 µm. **b**, Color map illustrating the lasing wavelengths for a region of 100 nanodisks on the sample shown in **a**, showing an increase in wavelength with increasing disk diameter. **c**, Log-log plot of light intensity emitted by a detached nanodisk laser in cell medium as a function of pump intensity (black symbols) and fit to the data with rate-equation model (red line). **d**, Emission spectra for same nanodisk and range of pump intensities as in c. **e**, Spectral stability of lasing peak for a typical nanodisk in cell medium; intensity of pump pulses, 100 µJ/cm$^2$.

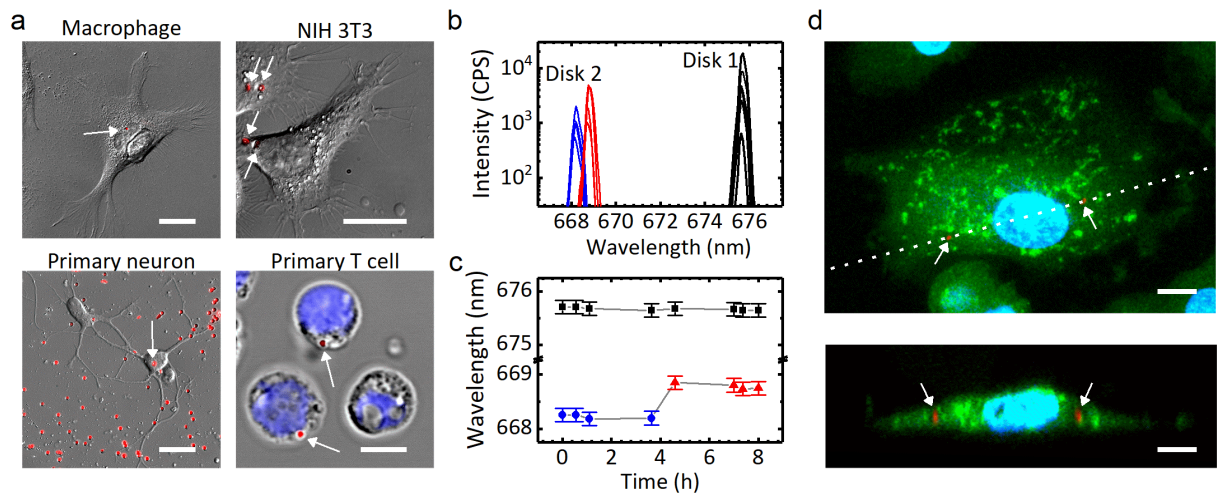

**Fig. 3 | Cellular uptake and lasing from semiconductor nanodisks. a**, DIC microscopy of macrophage, NIH 3T3, primary neuron and T cell with internalized nanodisks (overlaid red fluorescence, indicated by white arrows). Nucleus of T cell labelled by blue Hoechst dye. **b**, Laser spectra collected over period of 8 h for nanodisk inside a macrophage (Disk 1, black) and of a second disk that is internalized by the same cell during the experiment (Disk 2; blue before uptake, red after uptake). **c**, Peak wavelength for spectra in b over time. Error bars indicate FWHM of spectra. **d**, Laser scanning confocal fluorescence microscopy of fixed macrophage with nanodisks (red fluorescence, indicated by white arrows), nucleus in blue (Hoechst), and cytosol in green (Calcein). Maximum intensity projection (top) and vertical cross-section along the dotted line in the top panel. All scale bars, 20 µm; except for primary T cell, 5 µm.

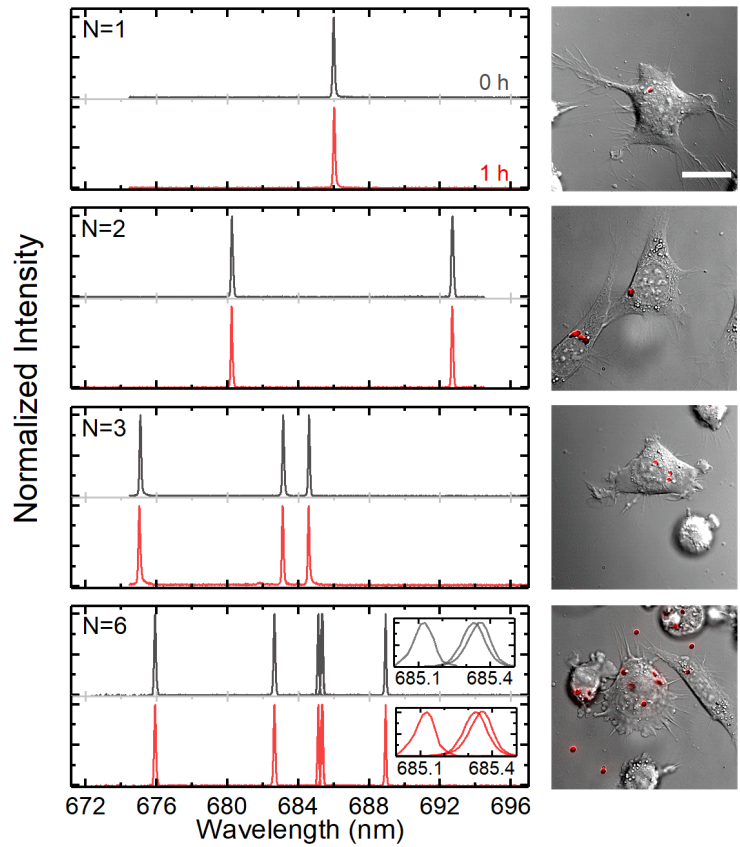

**Fig. 4 | Demonstration of optical barcoding of cells with multiple nanodisk lasers.** Emission spectrum from cells with N = 1, 2, 3, and 6 internalized nanodisk lasers, comparing spectra at the beginning of the experiment and after 1 h (left). DIC microscopy images of cells with overlaid red fluorescence from nanodisks (right). Inset for N = 6 shows spectra collected for individual excitation of three lasers with similar emission spectra. Scale bar, 20 µm.

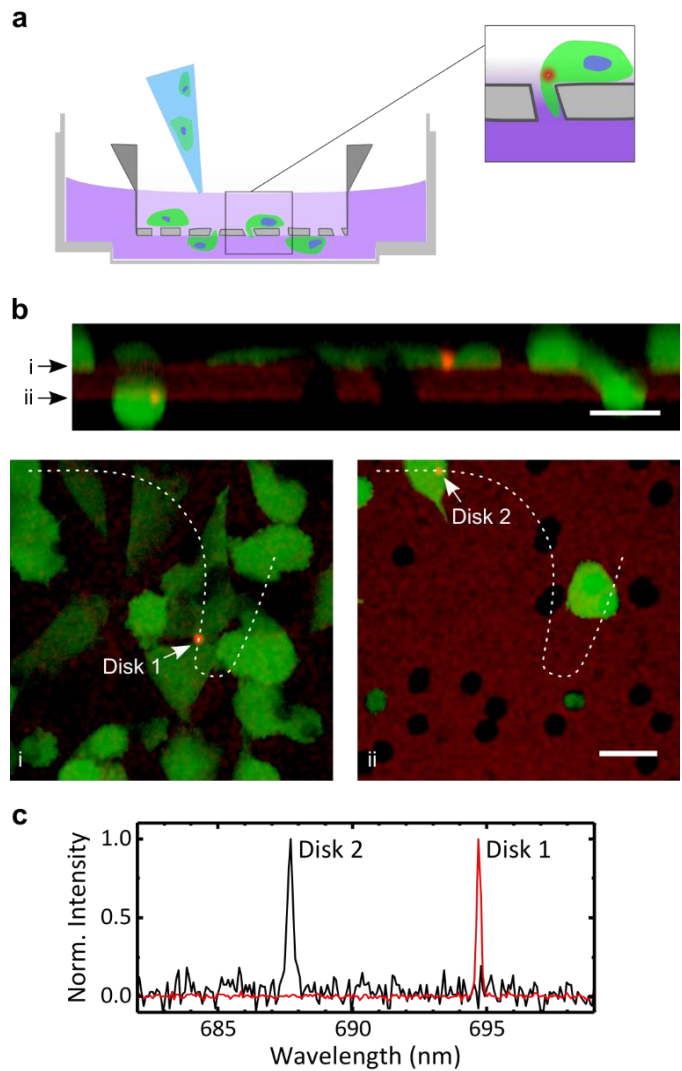

**Fig. 5 | Migration of cells with nanodisk lasers through microporous membrane. a**, Schematic cross-section of transwell migration assay, with cell culture dish, membrane insert and disk containing cells that are seeded into low-nutrient medium on the top side of membrane before migrating through pores in membrane towards nutrient-rich medium on the lower side. **b**, Live cell laser scanning confocal microscopy of membrane insert (weak red auto-fluorescence) with eGFP labelled cells (green) and nanodisk lasers (bright red). Arrows in top panel indicate the xy-slices shown in the bottom panels; dashed lines in bottom panels indicate the path of the cross-section in the top panel. Images use a logarithmic colour scale to visualize weak membrane fluorescence and bright disk fluorescence simultaneously. Scale bars, 20 µm. **c**, Lasing spectra of Disks 1 and 2 in b, recorded in parallel with the confocal microscopy.